\documentclass[12pt]{article}
\usepackage[a4paper,margin=1in]{geometry}
\usepackage[titletoc,title]{appendix}
\usepackage{changepage}
\usepackage[utf8x]{inputenc}
\usepackage{textcomp,marvosym}
\usepackage{fixltx2e}
\usepackage{amsmath,amssymb}
\usepackage{cite}
\usepackage{nameref,hyperref,nccmath}
\usepackage{authblk}
\usepackage{algorithm}
\usepackage{algpseudocode}
\usepackage{dsfont}
\usepackage{xcolor}
\usepackage{enumitem}
\usepackage{caption,setspace}
\usepackage{subcaption}
\usepackage{pdflscape}
\usepackage{graphicx}
\usepackage{placeins}
\usepackage{multirow,mathtools,nccmath}
\DeclareMathAlphabet\mathbfcal{OMS}{cmsy}{b}{n}
\def\cbl{\color{black}}
\def\cb{\color{black}}
\DeclarePairedDelimiter{\nint}\lfloor\rceil

\usepackage{mathtools}

\renewcommand{\eqref}[1]{Equation~(\ref{#1})}

\renewcommand{\baselinestretch}{1.5}

\def\cbl{\color{black}}
\def\cb{\color{black}}

\title{\sc{Discrete and continuous mathematical models of sharp--fronted collective cell migration and invasion}}

\author[1]{Matthew~J. Simpson\footnote{To whom correspondence should be addressed. E-mail: matthew.simpson@qut.edu.au}}
\author[1]{Keeley~M. Murphy}
\author[1]{Scott~W. McCue}
\author[1]{Pascal~R. Buenzli}

\affil[1]{School of Mathematical Sciences, Queensland University of Technology, Brisbane, Queensland 4001, Australia.}

\begin{document}

\maketitle

\begin{abstract}
Mathematical models describing the spatial spreading and invasion of populations of biological cells are often developed in a continuum modelling framework using reaction--diffusion equations.  While continuum models based on linear diffusion are routinely employed and known to capture key experimental observations, linear diffusion fails to predict well--defined sharp fronts that are often observed experimentally.  This observation has motivated the use of nonlinear degenerate diffusion, however these nonlinear models and the associated parameters lack a clear biological motivation and interpretation.  Here we take a different approach by developing  a stochastic discrete lattice--based model incorporating biologically--inspired mechanisms and then deriving the reaction--diffusion continuum limit. Inspired by experimental observations, agents in the simulation deposit extracellular material, that we call a \textit{substrate}, locally onto the lattice, and the motility of agents is taken to be proportional to the substrate density.  Discrete simulations that mimic a two--dimensional circular barrier assay illustrate how the discrete model supports both smooth and sharp--fronted density profiles depending on the rate of substrate deposition.  Coarse--graining the discrete model leads to a novel partial differential equation (PDE) model whose solution accurately approximates averaged data from the discrete model.  The new discrete model and PDE approximation provides a simple, biologically motivated framework for modelling the spreading, growth and invasion of cell populations with well--defined sharp fronts.  Open source Julia code to replicate all results in this work is available on \href{https://github.com/ProfMJSimpson/DiscreteSubstrate}{GitHub}.
\end{abstract}

\newpage
\section{Introduction}%
Continuum partial differential equation (PDE) models have been used for over 40 years to model and interpret the spatial spreading, growth and invasion of populations of cells~\cite{Balding1985,Sherratt90,Murray02}. \cbl PDE models have been used to improve our understanding of various biological processes including wound healing~\cite{Dale1994,sheardowncheng96,Savla04,Maini2004a,Maini2004b,Flegg2020}, embryonic development~\cite{Simpson2006,Simpson2007,Baker2008,Giniunaite2020}, tissue growth~\cite{Buenzli2020,Buenzli2022,Sengers2007}  as well as disease progression, such as cancer~\cite{Byrne2010,Swanson2003,Anderson1998,Haridas2017,Jarrett2018,Rockne2019,Hormuth2019,Murphy2024}.  \cb For a homogeneous population of cells with density $u \ge0 $, a typical PDE model can be written as
\begin{equation}\label{eq:General}
\dfrac{\partial u}{\partial t} = - \nabla \cdot \mathbfcal{J} + \mathcal{S},
\end{equation}
where $\mathbfcal{J}$ is the flux of cells and $\mathcal{S}$ is a source term that can be used to model proliferation and/or cell death. Different PDE models are specified by choosing different forms of $\mathbfcal{J}$ and $\mathcal{S}$.  Within the context of modelling homogenous cell populations, the most common choice for the flux term is based on the assumption that cells move randomly~\cite{Codling2008}, giving rise to linear diffusion with a flux term given by Fick's law, $\mathbfcal{J}  = -D \nabla u$, where $D > 0$ is the cell diffusivity~\cite{Murray02,Maini2004a,Maini2004b}.  A standard choice for the source term is to specify a logistic term to represent carrying capacity-limited proliferation, $\mathcal{S} = \lambda u (1 - u/K)$ where $\lambda>0$ is the proliferation rate and $K>0$ is the carrying capacity density~\cite{Murray02,Maini2004a,Maini2004b}.  These choices of $\mathbfcal{J}$ and $\mathcal{S}$ mean that Equation \ref{eq:General} is a multi-dimensional generalisation of the well-known Fisher-Kolmogorov model~\cite{Fisher1937,Kolmogorov1937,Canosa1973,ElHachem2019}, which has been successfully used to interpret a number of applications including \textit{in vivo} tumour progression~\cite{Swanson2003}, \textit{in vivo} embryonic development~\cite{Simpson2006}, \textit{in vitro} wound healing~\cite{Maini2004a,Maini2004b} and tissue growth~\cite{Buenzli2022,Sengers2007}.

\begin{figure}[H]
\centering
\includegraphics [width=1.0\textwidth]{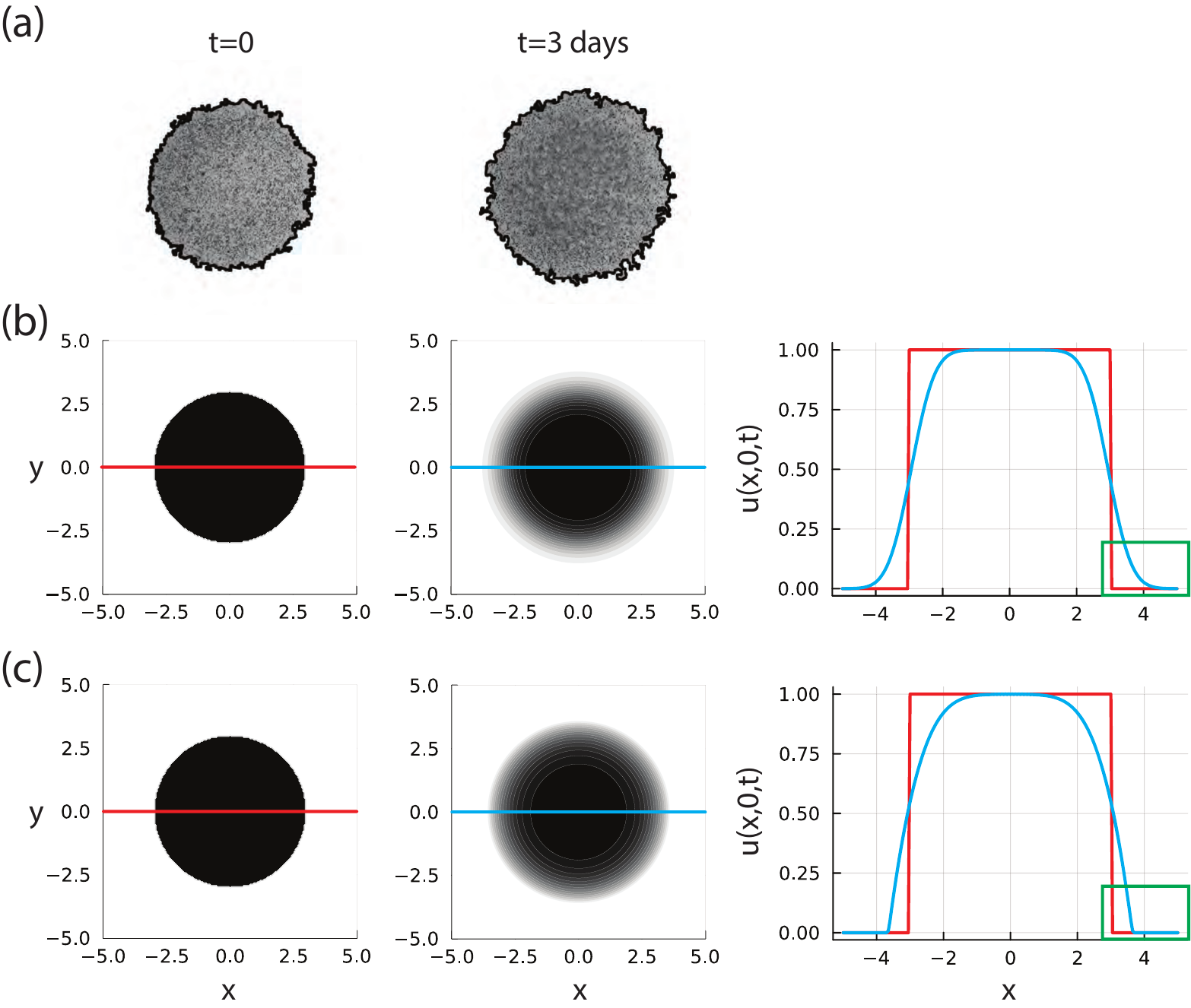}
\renewcommand{\baselinestretch}{1.0}
\caption{(a) Experimental images showing a population of non-proliferative fibroblast cells spreading in a two--dimensional barrier assay.  The image at $t=0$ shows the population just as the barrier is lifted, and the image at $t=3$ days showing the population of migrating cells spreading symmetrically with a sharp front.  Images reproduced from Simpson et al.~\cite{Simpson2013} with permission. (b)--(c) Numerical solutions of Equation \ref{eq:General} with $\mathbfcal{J} = -D \nabla u$ and $\mathbfcal{J} = -D u \nabla u$, respectively.  Both numerical solutions have $\mathcal{S}=0$ and $u(x,y,0)=1$ inside a disc of radius 3~mm, and $u(x,y,0)=0$ elsewhere to match the initial distribution of cells in the experiments shown in (a).  The numerical domain is a square of side length 10, and Equation \ref{eq:General} is discretised on a $201 \times 201$ uniform mesh.  The numerical solution of  Equation \ref{eq:General} at $t=3$ days is given in the middle panel of (b)--(c), and the details of the density profile are shown in the right-most panels where $u(x,0,t)$ is plotted at $t=0$ (red) and $t=3$ days (blue).  Details of the leading edge of the profiles are highlighted in the green rectangle near $x=4$ illustrating that the density profile in (b) has $u > 0$ at all locations, whereas the density profile in (c) has compact support, which is consistent with the experimental images in (a). All values of $x$ and $y$ in (b)--(c) measure location in terms of mm to be consistent with the experimental images in (a).  The numerical solution in (b) corresponds to a typical value of $D = D_{1} = 2100$~$\mu$m$^2$/hour for linear diffusion~\cite{Simpson2013}, and in (c) we set $D=D_{2}=4200$~$\mu$m$^2$/hour to satisfy $\int_{0}^{1} D_{1} \, \textrm{d}u = \int_{0}^{1}  D_{2} u \, \textrm{d}u$, to ensure that both the linear and nonlinear diffusion models lead to a similar amount of spreading over the experimental timescale~\cite{Landman1997}. } \label{F1}
\end{figure}

Figure \ref{F1}(a) shows experimental images of a simple two--dimensional \textit{in vitro} cell migration experiment, called a \textit{barrier assay}~\cite{Simpson2013,Das2015}.  These experiments are initiated by uniformly placing approximately 30,000 fibroblast cells as a monolayer inside a circular barrier of radius 3~mm.  In these experiments cells are pre-treated with an anti-mitotic drug that prevents proliferation~\cite{Sadeghi1998}, and there is no observed cell death~\cite{Simpson2013}.  Accordingly, we model this experiment by setting $\mathcal{S} = 0$ in Equation \ref{eq:General}.  The experiment proceeds by lifting the barrier at $t=0$ and observing how the population of cells spreads over time, with the right--most image in Figure \ref{F1}(a) showing the extent to which the population has spread after $t=3$ days.   Two key features of this experiment are immediately clear from these images: (i) the population of cells spreads symmetrically with time; and (ii) the experimental image at $t=3$ days shows a clear well-defined sharp front at the leading edge of the population as it spreads.  Images in Figure \ref{F1}(b) show a numerical solution of Equation \ref{eq:General} with $\mathcal{S}=0$ and the standard choice of linear diffusion, $\mathbfcal{J}  = -D \nabla u$, for a typical choice of $D$~\cite{Simpson2013}.  Consistent with the experiments in Figure \ref{F1}(a) we see that the simulated population spreads symmetrically, but plotting the density along the line $y=0$, in the right--most panel of Figure \ref{F1}(b) shows that we have $u > 0$ for all $x$ which is inconsistent with the well-defined sharp fronts at the leading edge in the experimental images.  This property of having $u > 0$ for all $x$ persists for all $t > 0$ which is a well--known deficiency of linear diffusion~\cite{Shigesada1997}.  Figure \ref{F1}(c) shows a numerical solution of Equation \ref{eq:General} with $\mathcal{S}=0$ and a nonlinear degenerate diffusive flux, $\mathbfcal{J}  = -D u \nabla u$, for a typical choice of $D$ in this model~\cite{Maini2004a,Maini2004b,Sengers2007,Sanchez1995,Jin2016,Simpson2024}.  Consistent with the experiments we see that the simulated population spreads symmetrically, and plotting the solution along the line $y=0$ in the right--most panel of Figure \ref{F1}(c) shows that we have a well--defined sharp front; $u > 0$ for $|x|<  X(t)$, and $u = 0$ for $|x| \ge X(t)$, where $X(t)$ is the front location at time $t$.    Full details of our numerical method for solving Equation \ref{eq:General} are given in the Appendix.

The qualitative comparison between the solution of the linear diffusion equation, the nonlinear degenerate diffusion equation and the experimental images in Figure \ref{F1} has been made with $\mathcal{S}=0$ so that the continuum PDE model is consistent with the experiments where proliferation is suppressed.  However, the difference between spreading cell fronts having sharp or smooth fronts is also relevant for models with $\mathcal{S} \ne 0$~\cite{Murray02,Maini2004a,Maini2004b}.   Throughout the first part of this work we set $\mathcal{S}= 0$, noting that the difference between smooth and sharp-fronted solutions of Equation \ref{eq:General} is, in general, determined by the choice of $\mathbfcal{J}$ rather than $\mathcal{S}$.  We will come back to this point in Section 2\ref{sec:proliferation} and provide evidence to support this claim.

\newpage
Many continuum models of homogeneous cell populations  adopt a simple linear diffusive flux, $\mathbfcal{J}  = -D \nabla u$, and this approximation is often made with the implicit or explicit acknowledgment that solutions of this PDE model fail to predict a well-defined front as observed experimentally.  In contrast, working with the degenerate nonlinear diffusion model by setting $\mathbfcal{J}  = -D u \nabla u$, can lead to a better match with experimental data with well--defined sharp fronts~\cite{Maini2004a,Maini2004b,Sengers2007,Jin2016,Warne2019,Harris04}.  With this choice of flux and $\mathbfcal{S}=0$, Equation \ref{eq:General} is also known as the \textit{porous medium equation}~\cite{Vazquez2006,Aronson1985,Barenblatt96,Pattle1959,Johnston2023}. Working with the degenerate nonlinear diffusion model is complicated by the fact that this model is one member of a family of models obtained by setting $\mathbfcal{J}  = -D u^n  \nabla u$, where $n>0$ is some constant.  Solutions of Equation \ref{eq:General} with this more general choice of nonlinear flux also leads to symmetric spreading with a well--defined sharp front like we saw in Figure \ref{F1}(c) for all values of $n > 0$.  These sharp--fronted solutions with compact support are similar to moving boundary problems in the sense that there is a well--defined front location with zero density, and the position of this front evolves with time which we can interpret as a model of the position of the cell front in an experiment~\cite{Vazquez2006,Aronson1985,Barenblatt96,Pattle1959,Johnston2023}.  The question of how to choose the value of the exponent $n$ remains unclear.  For example,  Sherratt and Murray~\cite{Sherratt90} studied an \textit{in vivo} wound healing experiment with $n = 0, 1$ and 4, and showed that all three choices of exponent could be used to make their reaction--diffusion PDE model match their experimental data.  Later, Jin et al~\cite{Jin2016}  studied a series of \textit{in vitro} scratch assays by setting $n =0,  0.5, 1, 2, 3$ and $4$ and concluded that $n = 1$ led to the best match to their experimental data without attempting to provide a biological motivation or interpretation of this choice of $n$.   Similarly, McCue et al.~\cite{McCue2019} studied a series of two--dimensional \textit{in vitro}  wound closure experiments and also found that $n=1$ provided the best match to their experimental data.  Other continuum modelling studies have simply worked with $n=1$ without explicitly considering other choices of the exponent~\cite{Buenzli2020,Sengers2007,Falco2023}.  In summary, a key challenge in using continuum PDE models with this generalised nonlinear degenerate diffusivity is that the exponent $n$ often acts as a fitting parameter~\cite{Warne2019}, and lacks clear a biological interpretation.  In addition to using these kinds of degenerate diffusion models to interpret biological observations, there is also a great deal of inherent mathematical interest in these models and their solutions~\cite{Sanchez1995,Sherratt1996}

An alternative to working with a continuum model to understand the collective spatial spreading, growth and invasion of cell populations is to work with a discrete modelling framework that considers the stochastic motion of individual cells~\cite{Anderson1998,Codling2008}.  \cbl Many kinds of discrete models of cell populations have been implemented to interpret experimental observations ranging from simple lattice-based models~\cite{Simpson2010,Mort2016} to more complicated lattice-free~\cite{Hoeme2010,Ghaffarizadeh2018} and vertex--based models~\cite{Smith2012,Mirams2013,Osborne2017,Crawshaw2023}. \cb  An attractive feature of working with discrete models is that experimental images and time--lapse movies showing individual cellular--level behaviours can be translated into a set of individual \textit{rules} that can be implemented with a stochastic framework to provide a high fidelity simulation--based model capturing  the key biological processes of interest~\cite{Volkening2015,Macfarlane2018}.  Discrete models can be implemented to visualise snapshots of the spreading population in a way that is directly analogous to performing and imaging an experiment to reveal the positions of individual cells within the population.  Another advantage of working with discrete stochastic models is that the discrete mechanism can be coarse-grained into an approximate continuum model, which means that we can encode different individual-level \textit{rules} into a simulation--based model, and then convert these rules into approximate continuum PDE models, \cbl and the solution of these coarse-grained models can be compared with averaged discrete data obtained by repeated simulation~\cite{Simpson2010,Deroulers09,Simpson09,Markham2013,Terragni2016,Pillay2017,Martinson2020,Nardini2020,Bruna2021,VandenHeuval2024}.   \cb  As described previously, there has been a great deal of effort devoted to understanding how different forms of continuum PDE models predict smooth or sharp--fronted solution profiles, however far less attention has been devoted to understanding what individual--level mechanisms lead to smooth or sharp fronts in discrete models of cell migration.

All mathematical models discussed so far are simple in the sense that they involve a single PDE or a single population of agents in a discrete framework that can be used to describe spreading of homogeneous population of cells.  Of course, there are many other more complicated models of collective cell spreading that can lead to sharp--fronted solution profiles.  These models include coupled reaction--diffusion models of multiple \cbl interacting cell populations~\cite{Painter2003} \cb as well as discrete models describing multiple populations~\cite{Johnston2017}.  Other families of mathematical models include models that describe cell migration that involves biased movement along chemical gradients, such as chemotaxis or haptotaxis~\cite{Perumpanani1999,Pettet2000}.  Here we will focus on more fundamental mathematical models of simple homogeneous populations composed of one cell type only, and we do not explicitly consider any biased migration mechanism, such as chemotaxis or haptotaxis.

In this work we propose a simple, biologically-motivated, lattice-based discrete model of collective cell migration and proliferation.  The discrete model explicitly models how individual cells in a two--dimensional \textit{in vitro} experiment produce a biological substrate (e.g. biological macromolecules, extracellular material) that is deposited onto the surface of the tissue culture plate~\cite{Buenzli2020,Lanaro2021,ElHachem2022}. Substrate is produced at a particular rate, and deposited locally by individuals within the simulated population.  Individual agents within the stochastic model undergo an unbiased random walk at a rate that is proportional to local substrate concentration, and   crowding effects are incorporated by ensuring that each lattice site can be occupied by no more than a single agent.  As we will demonstrate, this simple biologically--inspired mechanism allows us to simulate cell spreading experiments similar to those in Figure \ref{F1}(a).  Through simulation,  we first show that altering the rate of substrate deposition visually impacts the sharpness of the agent density front.  A deeper mathematical understanding of these observations is obtained by coarse-graining the discrete mechanism to give a novel PDE model whose solution describes the average behaviour of the stochastic model.   One way to interpret this new PDE model is that it naturally describes a linear diffusion mechanism at spatial locations well--behind the leading edge of the population, as well as a more complicated transport mechanisms at the leading edge of the spreading population that gives rise to sharp--fronted solution profiles consistent with experimental observations.  We show that averaged data from the discrete model can be very well approximated by numerical solutions of the  new continuum--limit PDE.  In particular, both the continuum and discrete models predict the formation of sharp-fronted density profiles.  A careful examination of the new continuum limit PDE model allows us to interpret how the different terms in the model lead to the formation of sharp, sometimes non-monotone fronts.  We conclude this study by incorporating a minimal model of cell proliferation into the discrete model, coarse--graining the proliferative discrete mechanism and comparing averaged data from the discrete model with proliferation to numerical solutions of the new PDE model.

\section{Results and Discussion}

\subsection{Stochastic model and simulations} \label{sec:discrete}
To account for crowding effects, we implement a lattice-based exclusion process where each lattice site can be either vacant or occupied by, at most, a single agent~\cite{Simpson2010,Deroulers09}.   From this point forward we will use the word \textit{agent} to refer to individuals within the simulated population and the word \textit{cell} to refer to individuals within an experimental population of biological cells.  For simplicity we implement the model on a two--dimensional square lattice with lattice spacing $\Delta$.  Each site is indexed by $(i,j)$, where $i,j \in \mathbb{Z}_+$, and each site has position $(x,y) = (i\Delta, j\Delta)$.  The lattice spacing is taken to be the size of a typical cell diameter~\cite{Simpson2010,Simpson2013}.  In any single realisation of the stochastic model the occupancy of each site $(i,j)$ is a binary variable $U_{i,j}$, with $U_{i,j} = 1$ if the site is occupied, and $U_{i,j}=0$ if the site is vacant.  Each site is also associated with a substrate concentration, which is a continuous function of time, $\bar{S}_{i,j}(t) \in [0, \bar{S}_{\textrm{max}}]$, where $\bar{S}_{\textrm{max}}$ is the maximum amount of substrate that can be accommodated at each lattice site.  For simplicity we write $S_{i,j}(t) \in [0,1]$, where $S_{i,j}(t) = \bar{S}_{i,j}(t)/\bar{S}_{\textrm{max}}$ is the non-dimensional substrate density.

A random sequential update method is used to advance the stochastic simulations through time.  If there are $N$ agents on the lattice, during the next time step of duration $\tau$, $N$ agents are selected independently, at random, one at a time with replacement, and given the opportunity to move.  If the chosen agent is at site $(i,j)$, the agent will attempt to move with probability $P S_{i,j}$, where $P \in [0,1]$ is the probability that an isolated agent will attempt to move during a time interval of duration $\tau$.  The target site for all potential motility events is selected at random from one of the four nearest neighbour lattice sites, and the potential motility event will be successful if the target site is vacant~\cite{Simpson2010,Simpson2013}.  Once $N$ potential movement events have been attempted, the density of substrate is updated by assuming that agents deposit substrate at a rate of $\Gamma$ per time step, so that the amount of substrate at each occupied lattice site increased by an amount $\Gamma \tau$, taking care to ensure that the maximum non-dimensional substrate density at each site is one.  In addition to specifying initial conditions for the distribution of agents and the initial density of substrate, we must specify values of two parameters to implement the stochastic simulation algorithm: $P \in [0,1]$ which determines the motility of agents, and $\Gamma > 0$ which determines the rate of substrate deposition.  With this framework a typical cell diffusivity is given by $D = P \Delta^2 / (4 \tau)$~\cite{Simpson2010}.

\begin{figure}[H]
\centering
\includegraphics [width=0.9\textwidth]{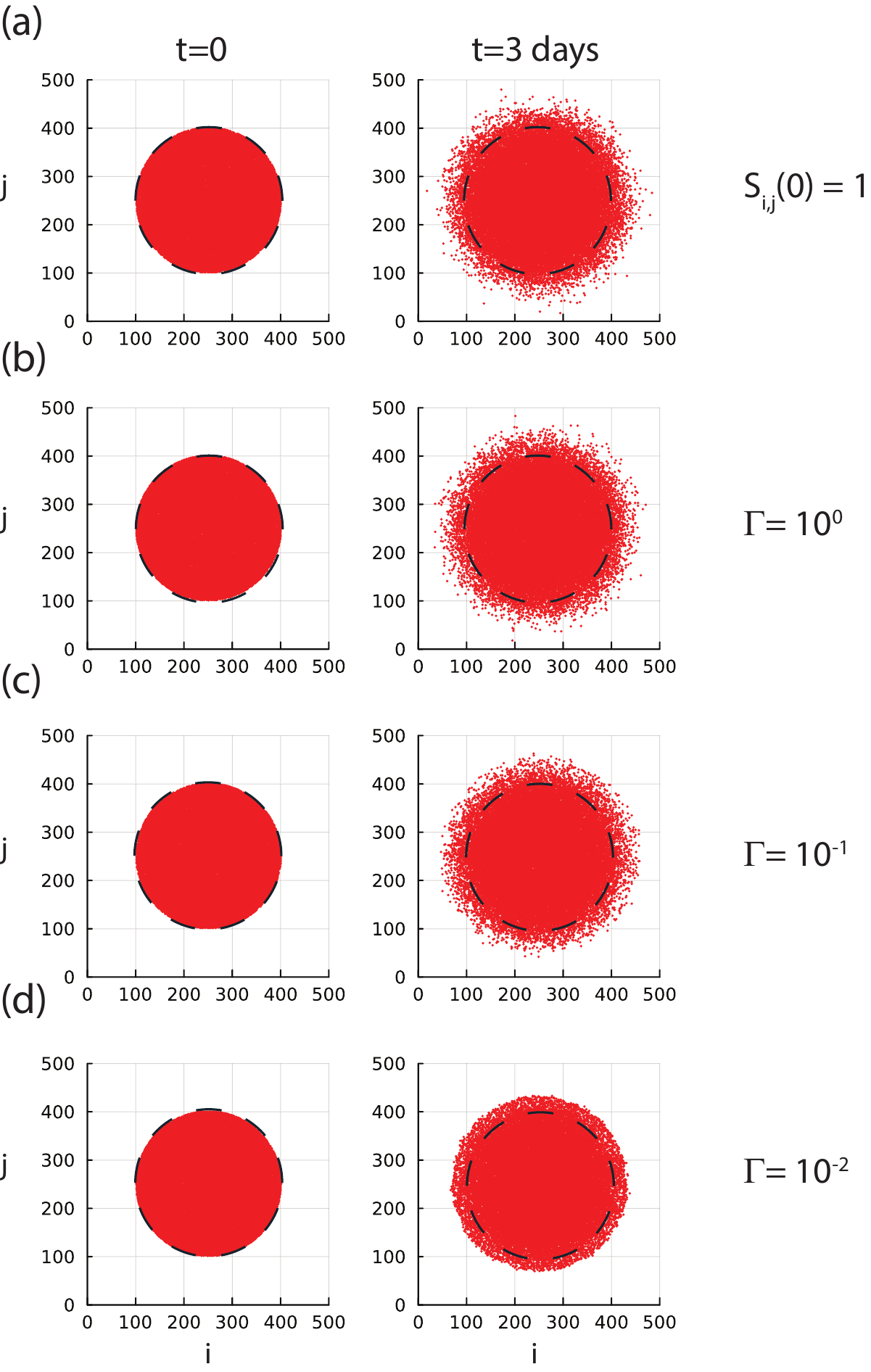}
\renewcommand{\baselinestretch}{1.0}
\caption{Discrete simulations illustrating the role of the substrate deposition rate $\Gamma$.  All simulations are performed on a $500 \times 500$ square lattice where the lattice spacing corresponds to 20 $\mu$m making the diameter of the simulated population distribution in the left column equal to the diameter of the barrier assay at $t=0$ in Figure \ref{F1}(a).  Simulations are initiated by randomly occupying sites within a circular region of radius 150 lattice sites so that the expected number of agents at the beginning of the simulation is 30,000.  Simulations are performed by setting $P=1$ and $\tau = 24/500$ hours, with values of $\Gamma$ as indicated.  Results in (a) correspond to initialising $S_{i,j}(0)=1$ at all lattice sites, whereas results in (b)--(d) correspond to initialising $S_{i,j}(0)=0$ at all lattice sites.  Each day of simulation corresponds to 500 time steps in the discrete model, and snapshots are reported in terms of the $(i,j)$ index of the lattice, which can be re-scaled to give the dimensional coordinates noting that $(x,y) = (i \Delta, j \Delta)$. \cbl Each subfigure shows a dashed line indicating the initial placement of the circular barrier.  Comparing the extent of the spreading populations with this dashed line gives a visual indication of the extent of spreading. \cbl} \label{F2}
\end{figure}

To illustrate how the discrete model can be used to model the barrier assay in Figure \ref{F1}(a) we perform a suite of simulations summarised in Figure \ref{F2}.  The radius of the barrier assay is 3~mm, and a typical cell diameter is approximately 20 $\mu$m~\cite{Simpson2010,Simpson2013}.  This  means we can simulate the initial placement of cells within the barrier by taking a circular region of radius $3000/20 =  150$ lattice sites to represent the disc enclosed by the barrier.  The experiments in Figure \ref{F1}(a) are initiated by placing approximately 30,000 cells uniformly, as a monolayer, within the circular barrier.  In the simulations we have $\nint{\pi 150^2} = 70,686$ lattice sites within the simulated barrier, and we initialise the simulations by randomly populating each lattice site within the barrier with probability $30,000 / 70,686 \approx 0.42$.  With the discrete model we can simulate a population of cells with a typical cell diffusivity of $D = 2100$ $\mu$m$^2$/hour~\cite{Simpson2013} by choosing $P=1$ and $\tau = 24/500$ hour.  This means that simulating 500 time steps of duration $\tau = 24/500$ hours is equivalent to one day in the experiment.  Results in Figure \ref{F2}(a) show a preliminary simulation with this initial condition where we set $S_{i,j}(0)=1$ at all lattice sites at the beginning of the simulation.  This first simulation corresponds to the simplest possible case where all lattice sites have the maximum amount of substrate present at the beginning of the experiment which means that the simulation does not depend upon the rate of deposition, $\Gamma$.  In Figure \ref{F2}(a) we see that the population of agents spreads symmetrically, and after 3 days we have a symmetric distribution of individuals without any clear front at the leading edge of the population.  In fact, by $t=3$ days we see that some agents within the simulated population become completely isolated, having spread far away from the bulk of the population as a result of chance alone.  This situation is inconsistent with the experimental images in Figure \ref{F1}(a) where we see a clear front at the leading edge of the population and a complete absence of individuals that become separated from the bulk population.  Open source Julia code to replicate these stochastic simulations is available on \href{https://github.com/ProfMJSimpson/DiscreteSubstrate}{GitHub}

Additional simulation results in Figure \ref{F2}(b)--(d) involve setting up the same initial distribution of agents as in Figure \ref{F2}(a) except that we set $S_{i,j}(0)=0$ at all lattice sites at the beginning of the simulation.  These simulations in Figure \ref{F2}(b)--(d) are more biologically realistic than the simulations in Figure \ref{F2}(a) because in the real experiment cells are placed into the barriers at the beginning of the experiment without having had any chance to deposit significant amounts of substrate onto the surface of the tissue culture plate before the barrier is lifted.  Simulations in Figure \ref{F2}(b)--(d) are shown for different substrate deposition rates, $\Gamma$.  If the substrate is deposited sufficiently fast, as in Figure \ref{F2}(b), the distribution of individual agents at $t=3$ days is visually indistinguishable from the case in Figure \ref{F2}(a) where we do not observe a clear front in the spreading population.  As $\Gamma$ is reduced, results in Figure \ref{F2}(c)--(d) show that the populations spread symmetrically with time, and now we see an increasingly well--defined sharp front as the population of agents spreads.  The snapshot of individuals in Figure \ref{F2}(d) shows that after $t=3$ days there are very few individual agents that are isolated away from the bulk of the population, and this distribution is consistent with the experimental observations in Figure \ref{F1}(a).

\subsection{Continuum limit partial differential equation model}\label{sec:pde}
We now provide greater mathematical understanding and interpretation of the discrete simulation results in Figure \ref{F2} by coarse-graining the discrete mechanism to give an approximate continuum limit description in terms of a PDE model in the form of Equation \ref{eq:General}.  We begin by considering the average occupancy of site $(i,j)$, where the average is constructed by considering a suite of $M$ identically--prepared simulations to give
\begin{equation}
\langle U_{i,j}\rangle(t) = \dfrac{1}{M} \sum_{m = 1}^{M}U_{i,j}^m(t),
\end{equation}
where $U_{i,j}^m(t)$ is the binary occupancy of lattice site $(i,j)$ at time $t$ in the $m$th identically--prepared realisation. With this definition we treat $\langle U_{i,j}\rangle(t) \in[0,1]$ as a smooth function of time, and for notational convenience we will simply refer to this quantity as $\langle U_{i,j}\rangle$.   Under these conditions we can write down an approximate conservation statement describing the change in average occupancy of site $(i,j)$ during the time interval from time $t$ to time $t + \tau$~\cite{Simpson2010,Deroulers09},

\begin{align}
\delta \langle U_{i,j} \rangle =& \dfrac{P}{4} \left[ \underbrace{\left(1 - \langle U_{i,j} \rangle \right) \sum \left[ S_{i,j} \langle U_{i,j} \rangle \right]}_{\textrm{migration onto site $(i,j)$}} - \underbrace{S_{i,j} \langle U_{i,j}  \rangle \left(4 -  \sum  \langle U_{i,j} \rangle  \right)}_{\textrm{migration out of site $(i,j)$}} \right], \label{eq:discreteu} \\
\delta S_{i,j} =& \begin{cases}
                 \Gamma \langle U_{i,j} \rangle  \quad \textrm{for} \quad S_{i,j} < 1,\\
                 0 \, \, \, \, \quad \textrm{for}  \quad S_{i,j} = 1, \label{eq:discretes}
                 \end{cases}
\end{align}
where, for notational convenience, we write
\begin{align}
\sum \left[ S_{i,j} \langle U_{i,j} \rangle \right] &= S_{i+1,j} \langle U_{i+1,j} \rangle + S_{i-1,j} \langle U_{i-1,j} \rangle + S_{i,j+1} \langle U_{i,j+1} \rangle + S_{i,j-1} \langle U_{i,j-1} \rangle, \\
\sum  \langle U_{i,j} \rangle &= \langle U_{i+1,j} \rangle +  \langle U_{i-1,j} \rangle +  \langle U_{i,j+1} \rangle +  \langle U_{i,j-1} \rangle.
\end{align}

The first term on the right of Equation \ref{eq:discreteu} approximately describes the increase in expected occupancy of site $(i,j)$ owing to motility events that would place agents on that site.  Similarly, the second term on the right of Equation \ref{eq:discreteu} approximately describes the decrease in expected occupancy of site $(i,j)$ owing to motility events associated with agents leaving site $(i,j)$. We describe these terms as \textit{approximate} as we have invoked the mean field assumption that the average occupancy of lattice sites are independent~\cite{Deroulers09}.  While this assumption is clearly questionable for any particular realisation of a discrete model, when we consider the expected behaviour of an ensemble of identically--prepared simulations this approximation turns out to be quite accurate~\cite{Simpson2010,Deroulers09}.  Note that setting $S_{i,j} = 1$ at all lattice sites means that this conservation statement simplifies to previous discrete conservation statements that have neglected the role of the substrate~\cite{Simpson2010}.

To proceed to the continuum limit we identify $\langle U_{i,j} \rangle$ and $S_{i,j}$ with smooth functions $u(x,y,t)$ and $s(x,y,t)$, respectively.   Throughout this work we associate uppercase variables with the stochastic model and lowercase variables with the continuum limit model.   We expand all terms in \ref{eq:discreteu} in a Taylor series about $(x,y) = (i\Delta, j\Delta)$, neglecting terms of $\mathcal{O}(\Delta^3)$ and smaller.  Dividing the resulting expressions by $\tau$, we take limits as $\Delta \to 0$ and $\tau \to 0$, with the ratio $\Delta^2/\tau$ held constant~\cite{Codling2008} to give
\begin{align}
\dfrac{\partial u}{\partial t} =& D \nabla \cdot \left[ s \nabla u + u\left(1-u\right) \nabla s \right], \label{eq:continuousu}\\
\dfrac{\partial s}{\partial t} =& \begin{cases}
                 \gamma u  \quad \textrm{for} \quad s < 1, \label{eq:continuouss} \\
                 0  \, \, \, \, \quad \textrm{for}  \quad s  = 1,
                 \end{cases}
\end{align}
where
\begin{equation}
D  = \lim_{\substack{\Delta \to 0 \\ \tau \to 0}} \left( \dfrac{P \Delta^2}{4 \tau} \right), \quad \gamma = \lim_{\substack{\Delta \to 0 \\ \tau \to 0}} \left(\dfrac{\Gamma}{\tau}\right),
\end{equation}
which relates parameters in the discrete model: $\Delta, \tau, P$ and $\Gamma$, to parameters in the continuum model: $D$ and $\gamma$.

The evolution equation for $s$, Equation \ref{eq:continuouss}, arises directly from our discrete model where we assume that each lattice site can occupy a maximum amount of substrate.  This leads to a mechanism that is very similar to an approach that has been recently adopted to study a generalisation of the well--known Fisher-KPP model where the nonlinear logistic source term is replaced with a linear \textit{saturation} mechanism~\cite{Berestycki2018a,Berestycki2018b,Berestycki2019}.  Solutions of these saturation--type models of invasion involve moving boundaries that form as a result of the saturation mechanism since this provides a natural moving boundary between regions where $s=1$ and $s < 1$. Later in Section 2\ref{sec:comparison} and 2\ref{sec:front} we will show that Equations  \ref{eq:continuousu}--\ref{eq:continuouss} can also be interpreted as moving boundary problem in exactly the same way as~\cite{Berestycki2018a,Berestycki2018b,Berestycki2019}.

The form of Equations \ref{eq:continuousu}--\ref{eq:continuouss} provides insight into the population--level mechanisms encoded with the discrete model.  To see this we write the flux encoded within Equation \ref{eq:continuousu} as,
\begin{equation}\label{eq:flux}
\mathbfcal{J} = \underbrace{-D s \nabla u}_{\text{diffusive flux}} - \underbrace{D u(1-u) \nabla s}_{\text{advective flux}}.
\end{equation}
Written in this way we can now interpret how these two components of the cell flux impact the population-level outcomes.  One way to interpret these terms is to note that the first term on the right of Equation \ref{eq:flux} is proportional to $\nabla u$ which is similar to a diffusive flux, and the second term on the right of Equation \ref{eq:flux} is proportional to $u(1-u)$ which acts like a non-linear advective flux.  In particular, this non-linear advective flux is similar to fluxes often encountered in mathematical models of traffic flow~\cite{Whitham1942}.

We can also interpret how the two components of $\mathbfcal{J}$ in Equation \ref{eq:flux} give rise to different features in the solution of the model depending on the location within a spreading population of individuals, such as the discrete populations shown in Figure \ref{F2}. For example, in regions that have been occupied by agents for a sufficiently long period of time, such as regions near the centre of the spreading populations in Figure \ref{F2}  where $u > 0$, locally we will eventually have $s = 1$ and $\nabla s = \mathbf{0}$.   This means that Equations \ref{eq:continuousu}--\ref{eq:continuouss} simplifies to the linear diffusion equation since the nonlinear advective flux vanishes and the diffusive-like flux simplifies to Fick's law of diffusion.  In contrast, regions that have been recently occupied by agents, such as near the leading edge of a population, we have $s < 1$ and $\nabla s \ne \mathbf{0}$.  Under these conditions  the diffusive flux is similar to a nonlinear diffusion term where the diffusive flux of $u$ is proportional to $s$, which reflects the fact that agent motility in the discrete model is directly proportional to the local density of substrate.  The advective-like component of the flux acts like a nonlinear advection term since the flux is proportional to $u(1-u)$~\cite{Whitham1942}, meaning that the advective flux vanishes when $u=0$ and $u=1$, and is a maximum when $u=1/2$.  The direction of the nonlinear advective flux  is opposite to $\nabla s$.  The nonlinear advective flux explicitly includes crowding effects encoded into the discrete model by enforcing that each lattice site can be occupied by, at most, a single agent.

\newpage
\subsection{Continuum--discrete comparison}\label{sec:comparison}
\begin{figure}[H]
\centering
\includegraphics [width=1.0\textwidth]{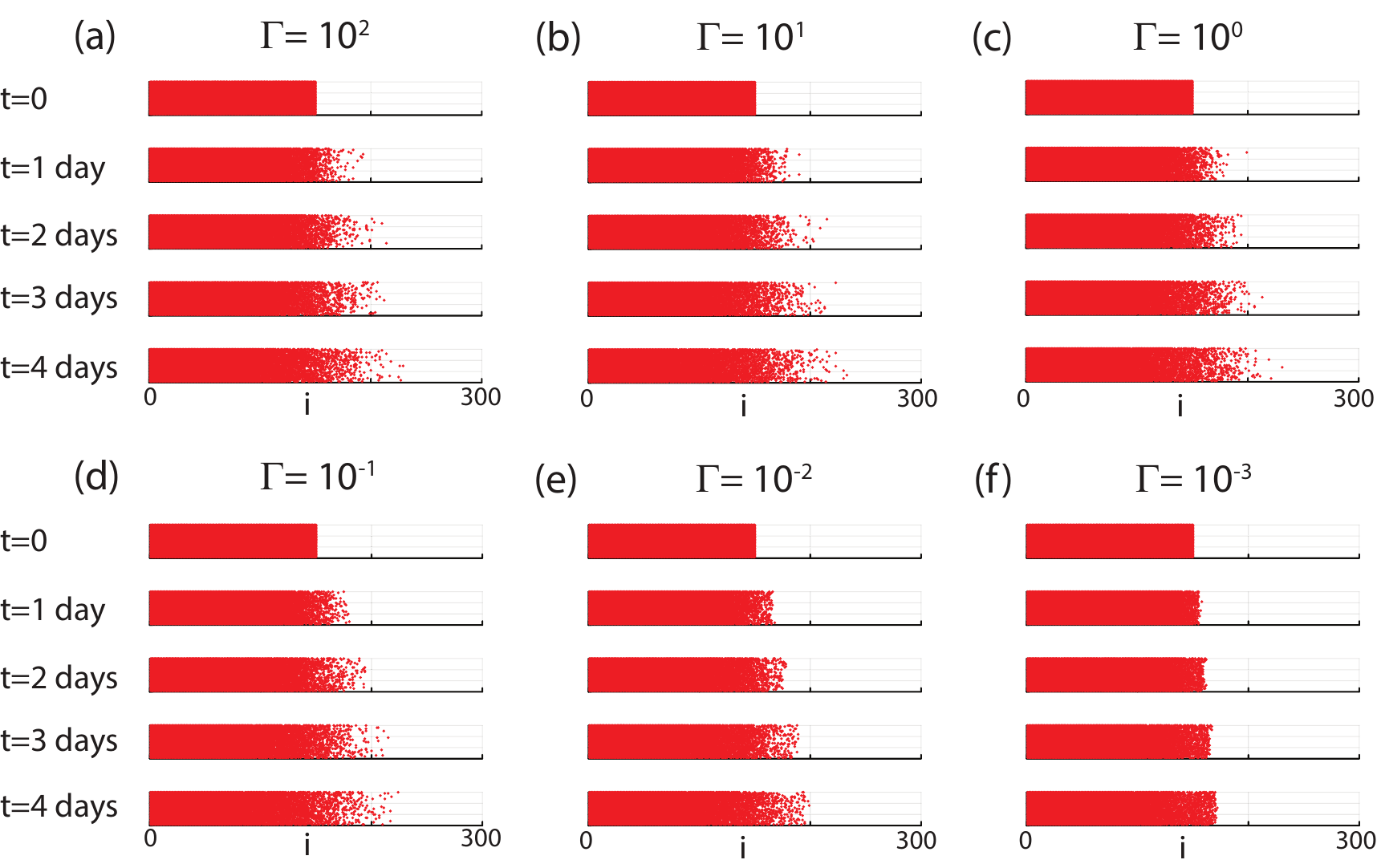}
\renewcommand{\baselinestretch}{1.0}
\caption{Discrete simulations with $P =1$, $\tau = 24/500$ hours, and various values of  $\Gamma$, as indicated.  All simulations are performed on a rectangular lattice of width $W=300$ and height $H=20$.  Simulations are initialised by setting $S_{i,j}=0$ at all lattice sites, and all sites with $i \le 150$ are occupied by agents.  Snapshots are shown at $t=1,2,3$ and $4$ days.  Each day of simulation corresponds to 500 time steps in the discrete model, and snapshots are reported in terms of the $(i,j)$ index of the lattice, which can be re-scaled to give the dimensional coordinates noting that $(x,y) = (i \Delta, j \Delta)$.} \label{F3}
\end{figure}

We
now  examine how well the numerical solution of Equations \ref{eq:continuousu}--\ref{eq:continuouss} matches averaged data from the discrete model.    The experimental images and stochastic simulations in Figures \ref{F1}--\ref{F2} correspond to a radially--symmetric polar coordinate system, which can be described by writing Equations \ref{eq:continuousu}--\ref{eq:continuouss} in terms of a radial coordinate system.  Instead, we consider a second set of discrete simulations, shown in Figure \ref{F3}, where we consider a rectangular domain with a width of 300 lattice sites, and height of 20 lattice sites.  Simulations are initialised by setting $S_{i,j}(0)=0$ at all lattice sites, and uniformly occupying all sites within  $i \le 150$ with agents.  \cbl Reflecting boundary conditions are imposed along all boundaries to ensure that the distribution of agents remains, on average, independent of vertical position~\cite{Simpson09,Simpson2010}, \cb and simulations are performed for $\Gamma = 10^{2}, 10^{1}, 10^{0}, 10^{-1}$ and $10^{-2}$ per time step, as shown in Figure \ref{F3}.  Simulation results are consistent with previous results in Figure \ref{F2} where we see that simulations performed with sufficiently large substrate deposition rates leads population spreading with a smooth front, without any obvious well--defined front position.  Simulations with larger $\Gamma$ lead to population spreading with a visually noticeable defined front.  Our main motivation for performing simulations on a rectangular--shaped lattice is that we can work with Equations \ref{eq:continuousu}--\ref{eq:continuouss} in a one--dimensional Cartesian coordinate system~\cite{Simpson2010}.

\newpage
Averaged agent density data are extracted from the simulations illustrated in Figure \ref{F3} by considering $M$ identically prepared realisations of the discrete model, averaging the occupancy of each lattice site across these realisations and then further averaging the occupancy along each column of the lattice to give~\cite{Simpson2010},
\begin{equation}\label{eq:1Dverticalaverage}
\langle U_i \rangle = \dfrac{1}{HM} \sum_{m=1}^{M}\sum_{j=1}^{H} U_{i,j}^m,
\end{equation}
where $H$ is the height of the lattice.  Numerical solutions of Equations \ref{eq:continuousu}--\ref{eq:continuouss} in a one--dimensional Cartesian coordinate system are obtained for parameter values and initial data consistent with the discrete simulations.  Details of the numerical method used to solve the continuum PDE model are given in the Appendix.  Results in Figure \ref{F4} compare numerical solutions of Equations \ref{eq:continuousu}--\ref{eq:continuouss} with averaged data from the discrete simulations, given by Equation \ref{eq:1Dverticalaverage} for various values of $\Gamma$, as indicated.

\begin{figure}[H]
\centering
\includegraphics [width=1.0\textwidth]{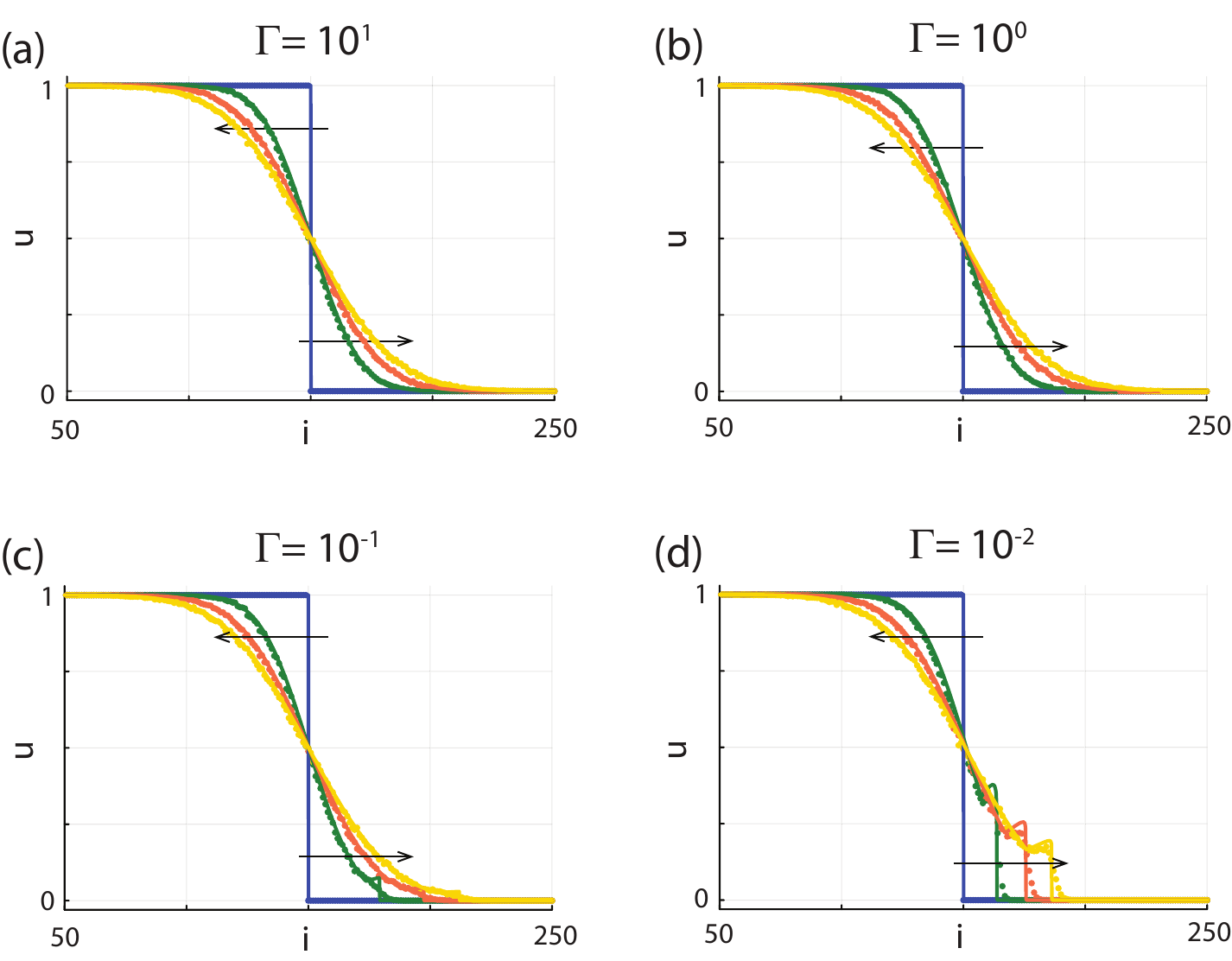}
\renewcommand{\baselinestretch}{1.0}
\caption{Averaged discrete data (dots) superimposed on numerical solutions of Equations \ref{eq:continuousu}--\ref{eq:continuouss} (solid).  Each subfigure compares averaged discrete data, constructed using Equation \ref{eq:1Dverticalaverage} with $H=20$ and $M = 100$, with a numerical solution of Equations \ref{eq:continuousu}--\ref{eq:continuouss}.  Four sets of solutions are shown for $\Gamma = 10^{1}, 10^{0}, 10^{-1}$ and $10^{-2}$ per time step, as indicated. Discrete simulations are initialised by occupying all lattice sites with $i \le 150$, and with $P =1$ and $\tau = 24/500$ hours. Within each subfigure a comparison is made at $t=0, 1, 2, 3$ and $4$ days shown in blue, green orange and yellow, respectively, as indicated.   Each day of simulation corresponds to 500 time steps in the discrete model, and snapshots are reported in terms of the $(i,j)$ index of the lattice, which can be re-scaled to give the dimensional coordinates noting that $(x,y) = (i \Delta, j \Delta)$.  The arrows within each subfigure show the direction of increasing time.} \label{F4}
\end{figure}

Results in Figure \ref{F4} indicate that the quality of the continuum--discrete match is very good for all values of $\Gamma$ considered.  For sufficiently large values of the substrate deposition rate in Figure \ref{F4}(a)--(b) we see that the density profiles are smooth, with no clear well--defined front location at the low density leading edge.  These results are consistent with the preliminary numerical results in Figure \ref{F1}(a)-(b) for the barrier assay geometry. In contrast, for sufficiently small values of the substrate deposition rate, density profiles in Figure \ref{F4}(c)--(d) show that we have a well--defined sharp front at the leading edge of the spreading populations.  The density profiles in Figure \ref{F4}(d) indicate that the solution of  Equations \ref{eq:continuousu}--\ref{eq:continuouss} for $u(x,t)$ has  compact support, and the density profiles are non-monotone with a small dip in density just behind the leading edge.  Interestingly, we see the small dip in density behind the leading edge in the averaged discrete data.  This indicates that the continuum limit PDE model provides an accurate approximation of the average densities from the discrete simulations.  \cbl As far as we are aware this dip in density just behind the leading edge has not been measured experimentally. \cb

\newpage
\subsection{Front structure}\label{sec:front}
\begin{figure}[H]
\centering
\includegraphics [width=1.0\textwidth]{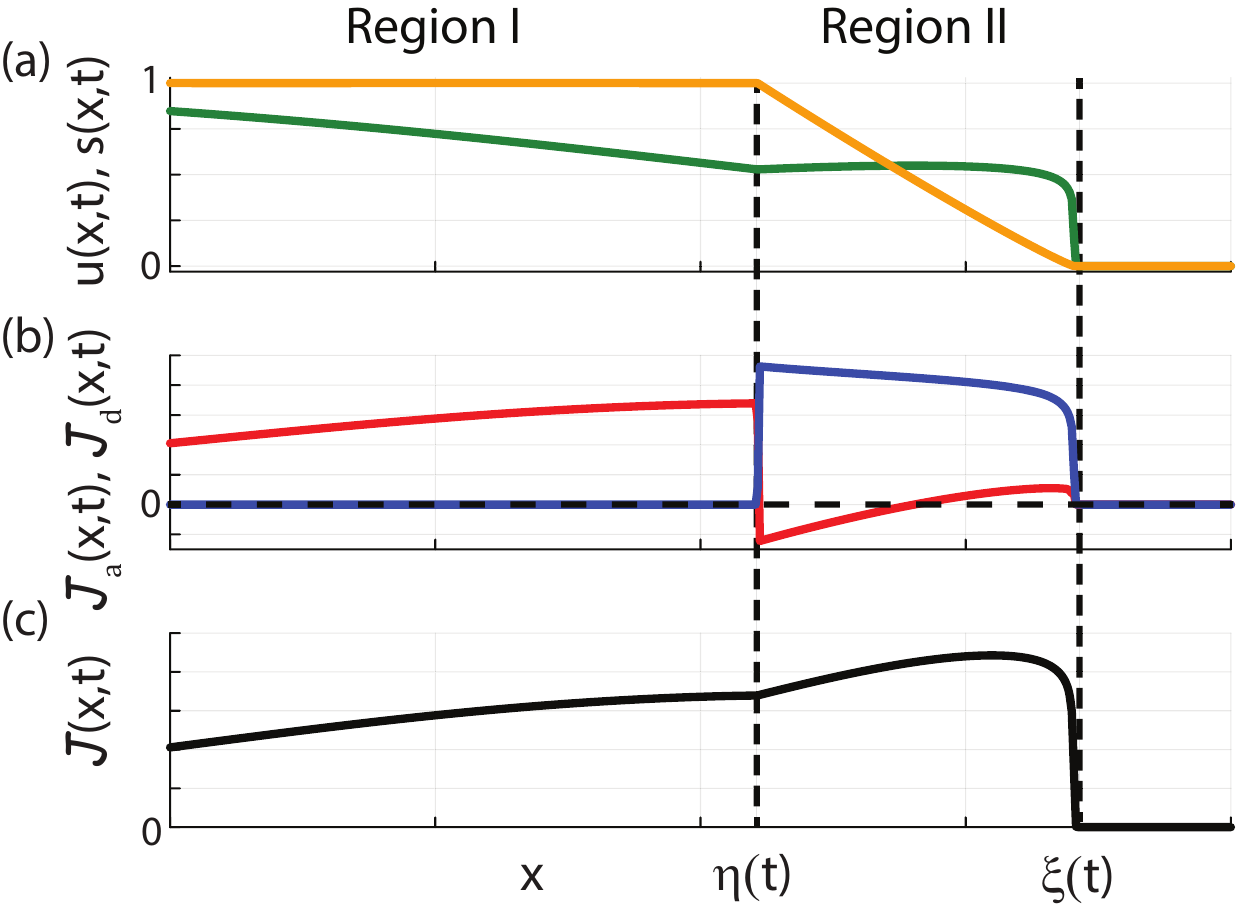}
\renewcommand{\baselinestretch}{1.0}
\caption{\cbl Schematic front structure. (a) Profiles for $u(x,t)$ (green) and $s(x,t)$ (yellow) within Region 1 for $x < \eta(t)$ and Region 2 for $\eta(t) < x < \xi(t)$.  (b) Using the profiles for $u(x,t)$ and $s(x,t)$ in (a) we show the corresponding spatial distribution of $\mathbfcal{J}_d(x,t) = -Ds\partial u / \partial x$ (red) and $\mathbfcal{J}_a(x,t) = -Du(1-u)\partial s / \partial x$ (blue). (c) Using the profiles for $u(x,t)$ and $s(x,t)$ in (a) we show the spatial distribution of the total flux $\mathbfcal{J}(x,t) = \mathbfcal{J}_d(x,t) + \mathbfcal{J}_a(x,t)$ (black).  The locations of $x = \eta(t)$ and $x = \xi(t)$ are shown with vertical dashed lines with $\xi(t) > \eta(t)$.  In (b) we include a horizontal dashed line at $\mathbfcal{J} = 0$ to emphasise the point that the diffusive flux changes sign at $x = \eta(t)$. \cb  } \label{F5}
\end{figure}

Now that we have confirmed that averaged data from the discrete model can be approximated by numerical solutions of Equations \ref{eq:continuousu}--\ref{eq:continuouss}, we will briefly describe and summarise the general features of the front structure in a simple one--dimensional Cartesian geometry analogous to the results in Figure \ref{F4}.    This discussion of the front structure is relevant for initial conditions of the form $s(x,0) = 0$ for all $x$, and $u(x,0) = 1 - \textrm{H}(X)$, where $\textrm{H}$ is the usual Heaviside step function so that initially we have $u = 1$ for $x < X$ and $u=0$ for $x > X$.  In all cases considered we impose zero flux boundaries on $u(x,t)$ at both boundaries of the one--dimensional domain.   Figure \ref{F5} shows a typical solution of Equations \ref{eq:continuousu}--\ref{eq:continuouss}.  This schematic solution corresponds to the most interesting case with sufficiently small $\gamma$ that we see a clear sharp--fronted solutions, and both $\partial u /\partial x$ and $\partial s /\partial x$ are discontinuous at some moving location $x = \eta(t)$.  As discussed in Section 2\ref{sec:pde}, the moving boundary at $x = \eta(t)$ arises because of the saturation mechanism governing the dynamics of $s$ in Equation \ref{eq:continuouss}.  This kind of moving boundary problem has been previously studied in the case of a generalised Fisher-KPP model~\cite{Berestycki2018a,Berestycki2018b,Berestycki2019}, except that these previous investigations have not involved any discrete stochastic models, or any kind of coarse-graining to arrive at an approximate PDE model.

The schematic showing $u(x,t)$ and $s(x,t)$ in Figure \ref{F5}(a) motivates us to consider two regions within the solution:
\begin{itemize}
\item Region 1: $x < \eta(t)$ where $s(x,t) = 1$, and \\
\item Region 2: $\eta(t) < x < \xi(t)$, where $0 < s(x,t) < 1$.
\end{itemize}
Ahead of Region 2 where $x > \xi(t)$ we have $u = s = 0$, and so we consider $x = \xi(t)$ to be the \textit{front} of the solution.  In Region 1 we have $s(x,t) = 1$ and $\partial s / \partial x = 0$, which means that the evolution of $u(x,t)$ in Region 1 is governed by the linear diffusion equation and the flux of $u$ simplifies to $\mathbfcal{J} = -D \partial u / \partial x$.  This simplification explains why, for this initial condition, $u(x,t)$ is a monotonically decreasing function of $x$ within Region 1 because solutions of the linear diffusion equation obey a maximum principle~\cite{Protter67}.

Region 2 is characterised by having $s(x,t) < 1$ with $\partial s / \partial x < 0$.  The interface between Region 1 and Region 2 has $s(\eta(t),t) = 1$ and $u(\eta(t),t) = u^*$, for some value $0 < u^* < 1$.  Within Region 2 the flux of $u$ is given by $\mathbfcal{J} = -D s\partial u / \partial x - Du(1-u) \partial s /\partial x$.  The advective component of the flux, $- Du(1-u) \partial s /\partial x$, is directed in the positive $x$ direction, which means that the flux of $u$ entering Region 2 across the interface at $x = \eta(t)$ is partly advected in the positive $x$ direction due to the advective flux term that acts within Region 2 only. This additional advective flux in the positive $x$ direction within Region 2 explains why there can be a local minima in $u$ at $x = \eta(t)$.  The diffusive component of the flux in Region 2, $-Ds \partial u / \partial x$, can act in either the positive $x$--direction when $\partial u / \partial x < 0$ or in the negative $x$--direction when $\partial u/ \partial x > 0$. The schematic in Figure \ref{F5}(a) shows $u(x,t)$ and $s(x,t)$ across Regions 1 and 2.  Associated schematics in Figure \ref{F5}(b) shows $\mathbfcal{J}_d = -D s \partial u / \partial x$ and $\mathbfcal{J}_a = -D u(1-u) \partial s / \partial x$, and the schematic in Figure \ref{F5}(c) shows  $\mathbfcal{J} =  \mathbfcal{J}_d+ \mathbfcal{J}_a$ for the $u$ and $s$ profiles in Figure \ref{F5}(a).  These plots of the fluxes show that while the total flux  $\mathbfcal{J} > 0$ across both Regions 1 and 2, we see that $\mathbfcal{J}_a$ vanishes everywhere except within Region 2, and $\mathbfcal{J}_d > 0$ within Region 1, but $\mathbfcal{J}_d$ changes sign within Region 2 in this case.

Exploring numerical solutions of Equations  \ref{eq:continuousu}--\ref{eq:continuouss} indicates that the width of Region 2, $w(t) = \xi(t) - \eta(t)$ decreases with $\gamma$.  This is both intuitively reasonable, and consistent with the observations in Figure \ref{F2} regarding how the structure of the front appeared to vary with the deposition rate in the discrete model.  Numerical solutions of Equations  \ref{eq:continuousu}--\ref{eq:continuouss} indicate that as  $\gamma \to \infty$ we have $s(x,t) \to 1 - \textrm{H}(\eta(t))$ and $w(t) \to 0^+$.  Since the width of Region 2 vanishes for sufficiently large $\gamma$, the solution of Equations \ref{eq:continuousu}--\ref{eq:continuouss} can be accurately approximated by the solution of the linear diffusion equation, which is independent of $\gamma$.  Again, this outcome is consistent with the discrete simulations in Figure \ref{F2} where we observed that simulations with large deposition rates were visually indistinguishable from simulations where all lattice sites were initialised with the maximum substrate concentration where the continuum limit of the discrete model is the linear diffusion equation~\cite{Simpson2010}.

The schematic profiles of $u(x,t)$ and $s(x,t)$ in Figure \ref{F5} can also be interpreted in terms of the mechanisms acting in discrete model.  When agents within Region 2, close to the front, move in the positive $x$--direction to a lattice site that has never been previously occupied, that agent will experience $S_{i,j}=0$ at the new site.  This means that the agent will be stationary for a period of time until that agent deposits substrate, which means that  $S_{i,j}$ increases.  While there is empty space behind that agent, for example at site $(i-1,j)$  where it was previously located, the agent cannot easily move back until a sufficient amount of time has passed to build up the amount of substrate.  Therefore, Region 2 within the discrete model involves acts as a low--motility zone where agents become momentarily stationary until sufficient substrate is produced to enable the agents to continue to move.

\cbl In addition to presenting a physical interpretation of the front structure, both from the continuum and discrete point of view summarised here, we also attempted to examine the structure of the front more formally using an interior layer analysis by identifying $\gamma^{-1}$ as a small parameter in the system of governing equations.  Unfortunately this leads to a nonlinear partial differential equation as the $O(1)$ problem that determines the shape of the interior layer.  Since we are unable to solve the $O(1)$ problem we did not proceed any further with this approach.  \cb

\subsection{Proliferation}\label{sec:proliferation}
\begin{figure}[H]
\centering
\includegraphics [width=1.0\textwidth]{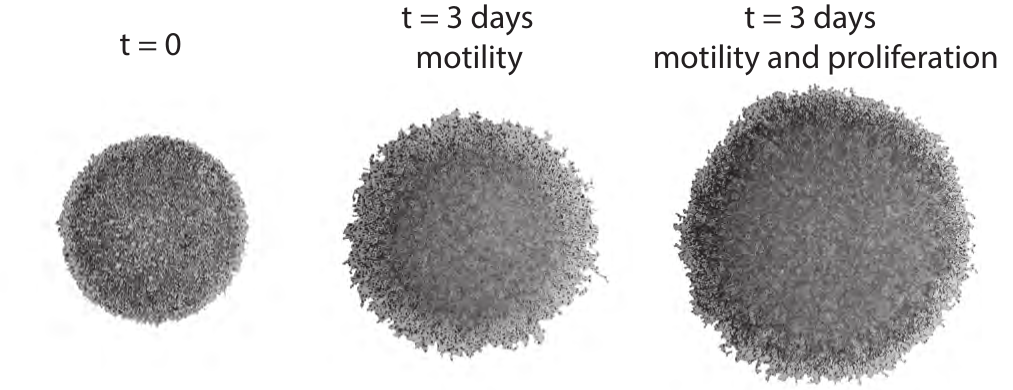}
\renewcommand{\baselinestretch}{1.0}
\caption{Circular barrier assay images comparing the spatial spreading of motile and non-proliferative population with the spatial spreading of a motile and proliferative population of fibroblast cells. The left--most image shows a barrier assay at $t=0$ days just after the circular barrier is lifted.  This experiment is initiated by placing approximately 30,000 fibroblast cells uniformly inside a barrier of radius 3~mm.  The central image shows the spatial extent of the population at $t=3$ days where the cell are pre-treated to suppress proliferation. The right--most image shows the spatial extent of the population at $t=3$ days where the cells are motile and proliferative.  All images reproduced from Simpson et al.~\cite{Simpson2013} with permission.} \label{F6}
\end{figure}

The experimental image in Figure \ref{F1}(a) shows a barrier assay describing the spatial spreading of a population of fibroblast cells that are pre-treated to prevent proliferation~\cite{Simpson2013,Sadeghi1998}.  All subsequent discrete and continuum modelling in this work so far has focused on conservative populations without any death or proliferation mechanisms so that these simulations are consistent with the preliminary experimental observations in Figure \ref{F1}.  In the discrete model this is achieved by simulating a population of $N$ agents, where $N$ is a constant.  In the continuum model this is achieved by working with PDE models like Equation \ref{eq:General} with $\mathcal{S} = 0$.  \cbl To conclude this study we now re-examine all discrete and continuum models by incorporating a minimal proliferation mechanism motivated by the additional experimental results summarised in Figure \ref{F6}.  The proliferation mechanism involves agent division only without any death process because there is no evidence of cell death in the experiments that motivate these simulations~\cite{Simpson2013}. \cb  The left--most image in Figure \ref{F6} shows a barrier assay initialised with approximately 30,000 fibroblast cells just after the barrier is lifted at $t=0$.  The central image in Figure \ref{F6} shows the outcome of a barrier assay where the fibroblast cells are pre-treated to suppress proliferation~\cite{Simpson2013,Sadeghi1998}, and the right--most image shows the outcome of a barrier assay that is initialised in the same way except that the fibroblast cells are not pre-treated to suppress proliferation.  This means that the right--most image in Figure \ref{F6} shows the outcome of a barrier assay in which fibroblast cells are free to move and proliferate~\cite{Simpson2013}.  The motile and proliferative population expands symmetrically, and the leading edge of the population remains sharp.  The main difference between the outcome of the barrier assays for the motile and proliferative population compared to the population where proliferation is suppressed is that cell proliferation leads to more rapid spatial expansion of the population.

\begin{figure}[H]
\centering
\includegraphics [width=1.0\textwidth]{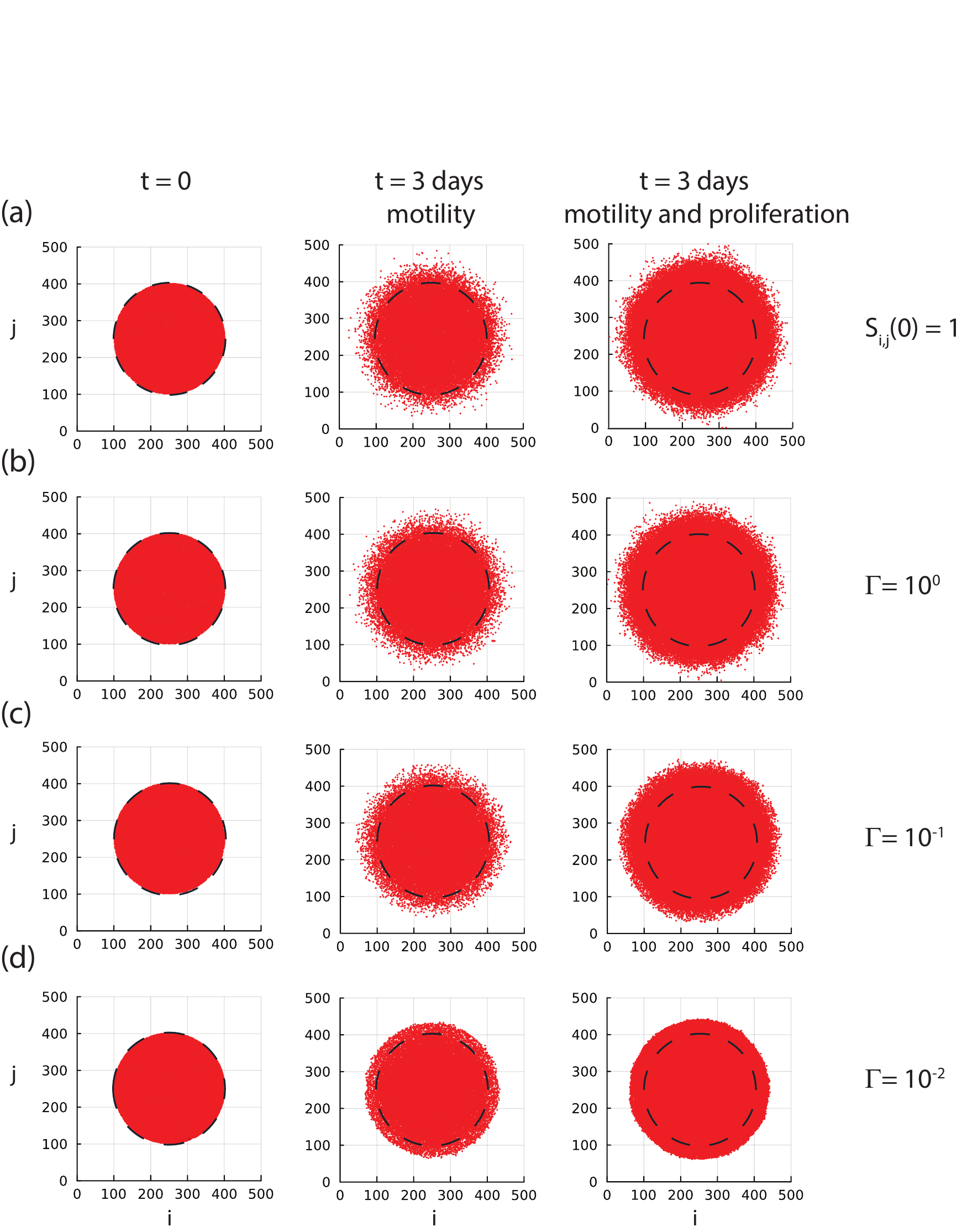}
\renewcommand{\baselinestretch}{1.0}
\caption{Discrete simulations illustrating the role of the substrate deposition rate $\Gamma$ in the spatial spreading of a population of motile agents without proliferation with the spatial spreading of a population of motile and proliferative agents.  All simulations are performed on a $500 \times 500$ square lattice where the lattice spacing corresponds to 20 $\mu$m making the diameter of the simulated populations in the left column equal to the diameter of the populations at $t=0$ in Figures \ref{F1} and \ref{F6}.
Simulations are initiated by randomly occupying sites within a circular region of radius 150 lattice sites so that the expected number of agents at the beginning of the simulation is 30,000.  Simulations of motile and non-proliferative populations correspond to $P=1$, $Q=0$ and $\tau = 24/500$ days, with values of $\Gamma$ as indicated.  Simulations of motile and proliferative populations correspond to $P=1$, $Q=1/500$ and $\tau = 24/500$ days, with values of $\Gamma$ as indicated.  Results in (a) correspond to initialising $S_{i,j}(0)=1$ at all lattice sites, whereas results in (b)--(d) correspond to initialising $S_{i,j}(0)=0$ at all lattice sites. Each day of simulation corresponds to 500 time steps in the discrete model, and snapshots are reported in terms of the $(i,j)$ index of the lattice, which can be re-scaled to give the dimensional coordinates noting that $(x,y) = (i \Delta, j \Delta)$. \cbl Each subfigure shows a dashed line indicating the initial placement of the circular barrier.  Comparing the extent of the spreading populations with this dashed line gives a visual indication of the extent of spreading. \cbl}\label{F7}
\end{figure}

A minimal model of proliferation is now incorporated into the discrete model described previously in Section 2\ref{sec:discrete}.  The key difference is that previously the number of agents $N$ remained fixed during the stochastic simulations, whereas now $N(t)$ is an non-decreasing function of time.  Within each time step of the discrete model, after giving $N(t)$ randomly--selected agents an opportunity to move, we then select another $N(t)$ agents at random, one at a time with replacement, and given the selected agents an opportunity to proliferate with probability $Q \in [0,1]$.  We take a simple approach and assume that the proliferation is independent of the local substrate density.  If a selected agent is going to attempt to proliferate, the target site for the placement of the daughter agent is randomly selected from one of the four nearest neighbour lattice sites~\cite{Callaghan06}.  If the target site is occupied then the proliferation event is aborted owing to crowding effects, whereas if the target site is vacant a new daughter agent is placed on the target site.  At the end of every time step we update $N(t)$ to reflect the change in total population owing to proliferation during that time step~\cite{Simpson2010,Callaghan06}.   A set of preliminary simulations comparing results with $Q=0$ and $Q > 0$ are given in Figure \ref{F7}.  In these simulations we compare the spatial spreading of 30,000 agents uniformly distributed within a circular region of diameter 3~mm.  Results in Figure \ref{F7}(a) are made in the case where we set $S_{i,j}(0)=1$ at all lattice sites at the beginning of the experiment, and we repeat the comparison for simulations with $S_{i,j}(0)= 0$ with $\Gamma = 10^{0}, 10^{-1}$ and $10^{-2}$ in Figure \ref{F7}(c)--(d), respectively.

Similar to the experiments in Figure \ref{F6}, our simulations in Figure \ref{F7} show that incorporating proliferation increases the rate at which the growing populations spread and invade the surrounding area.  As for the non-proliferative simulations in Figure \ref{F2} we see that the front of the spreading populations is poorly defined when $\Gamma$ is sufficiently large, with a relatively diffuse distrubution of agents that includes many isolated individuals that have migrated well--ahead of the bulk population.  In contrast, reducing $\Gamma$ leads to visually well--defined sharp fronts with a clearer boundary at the leading edge of the proliferative population.  These sharper fronts contain very few isolated individual agents.  Visual comparison of the proliferative and non-proliferative snapshots in Figure \ref{F7}(c)--(d) indicates that incorporating proliferation leads to an increasingly sharp and well--defined sharp front.  These simulations indicate that having substrate--dependent motility and substrate--independent proliferation is sufficient to produce sharp and well--defined fronts in the discrete simulations.

To interpret the differences between the motile populations and the motile and proliferative populations in Figure \ref{F7} we coarse grain the discrete model by following a similar approach taken in Section 2\ref{sec:pde}.   To proceed we write down an approximate conservation statement describing the change in average occupancy of site $(i,j)$ during the time interval from time $t$ to time $t + \tau$,
\begin{align}
\delta \langle U_{i,j} \rangle =& \dfrac{P}{4} \left[ \underbrace{\left(1 - \langle U_{i,j} \rangle \right) \sum \left[ S_{i,j} \langle U_{i,j} \rangle \right]}_{\textrm{migration onto site $(i,j)$}} - \underbrace{S_{i,j} \langle U_{i,j}  \rangle \left(4 -  \sum  \langle U_{i,j} \rangle\right)}_{\textrm{migration out of site $(i,j)$}} \right] \notag \\
+&\dfrac{Q}{4}\underbrace{\left(1 - \langle U_{i,j} \rangle \right) \sum  \langle U_{i,j} \rangle, }_{\textrm{proliferation onto site $(i,j)$}}   \label{eq:discreteu2} \\
\delta S_{i,j} =& \begin{cases}
                 \Gamma \langle U_{i,j} \rangle  \quad \textrm{for} \quad S_{i,j} < 1,\\
                 0 \, \, \, \, \quad \textrm{for}  \quad S_{i,j}  = 1. \label{eq:discretes2}
                 \end{cases}
\end{align}
The new term on the right of Equation \ref{eq:discreteu2} approximately describes the increase in expected density of site $(i,j)$ owing to proliferation events that would place an agent on that site provided that the target site is vacant~\cite{Simpson2010}.  To proceed to the continuum limit we again identify $\langle U_{i,j} \rangle$ and $S_{i,j}$ with smooth functions $u(x,y,t)$ and $s(x,y,t)$, respectively, and expand all terms in \ref{eq:discreteu} in a Taylor series about $(x,y) = (i\Delta, j\Delta)$, neglecting  terms of $\mathcal{O}(\Delta^3)$ and smaller.  Dividing the resulting expression by $\tau$, we take limits as $\Delta \to 0$ and $\tau \to 0$, with the ratio $\Delta^2/\tau$ held constant~\cite{Codling2008} to give
\begin{align}
\dfrac{\partial u}{\partial t} =& D \nabla \cdot \left[ s \nabla u + u\left(1-u\right) \nabla s \right] + \lambda u(1-u), \label{eq:continuousu2}\\
\dfrac{\partial s}{\partial t} =& \begin{cases}
                 \gamma u  \quad \textrm{for} \quad s < 1 \label{eq:continuouss2} \\
                 0  \, \, \, \, \quad \textrm{for}  \quad s   = 1,
                 \end{cases}
\end{align}
where
\begin{equation}\label{eq:ContinuumLimit2}
D  = \lim_{\substack{\Delta \to 0 \\ \tau \to 0}}  \left(\dfrac{P \Delta^2}{4 \tau}\right), \quad \lambda  = \lim_{\substack{\Delta \to 0 \\ \tau \to 0}} \left(  \dfrac{Q}{\tau} \right), \quad \gamma = \lim_{\substack{\Delta \to 0 \\ \tau \to 0}} \left(\dfrac{\Gamma}{\tau}\right),
\end{equation}
which provides relationships between parameters in the discrete model: $\Delta, \tau, P, Q$ and $\Gamma$, to parameters in the continuum model: $D$, $\lambda$, and $\gamma$.  The additional term in Equation \ref{eq:discreteu2} is simply a logistic source term with carrying capacity of unity, which reflects the fact that the occupancy of any lattice site is limited to a single agent.  The numerical method we use to solve Equations \ref{eq:continuousu2}--\ref{eq:continuouss2} is given in the Appendix.

It is straightforward to choose parameters to mimic known biological observations.  An important parameter for applying these models to biological experiments is the ratio $P/Q$, which compares the relative frequency of motility to proliferation events for isolated agents in regions where $S_{i,j}=1$.  Key parameters in an experiment are the cell diameter $\Delta$, the cell diffusivity $D$, and the proliferation rate $\lambda$, which is related to the cell doubling time, $t_{\textrm{d}}$, by $\lambda = \log_{\textrm{e}}2/t_{\textrm{d}}$.  Using Equation \ref{eq:ContinuumLimit2} we have $(Q/P) = \Delta^2 \log_{\textrm{e}} 2 /(4 D t_{\textrm{d}})$, noting that this ratio is independent of $\tau$.  With typical values of $\Delta=20$ $\mu$m, $D=2100$ $\mu$m$^2$/hour and $t_{\textrm{d}} = 16$ hours we have $Q/P  \approx 1/500$, which means setting $P=1$ and $Q=1/500$ correspond to biologically--relevant parameter values of the discrete model.  One way of interpreting this choice of parameters is that the average time between proliferation events for an isolated agent is 500 times longer than the average time between motility events in regions where $S_{i,j}=1$.

\begin{figure}[H]
\centering
\includegraphics [width=1.0\textwidth]{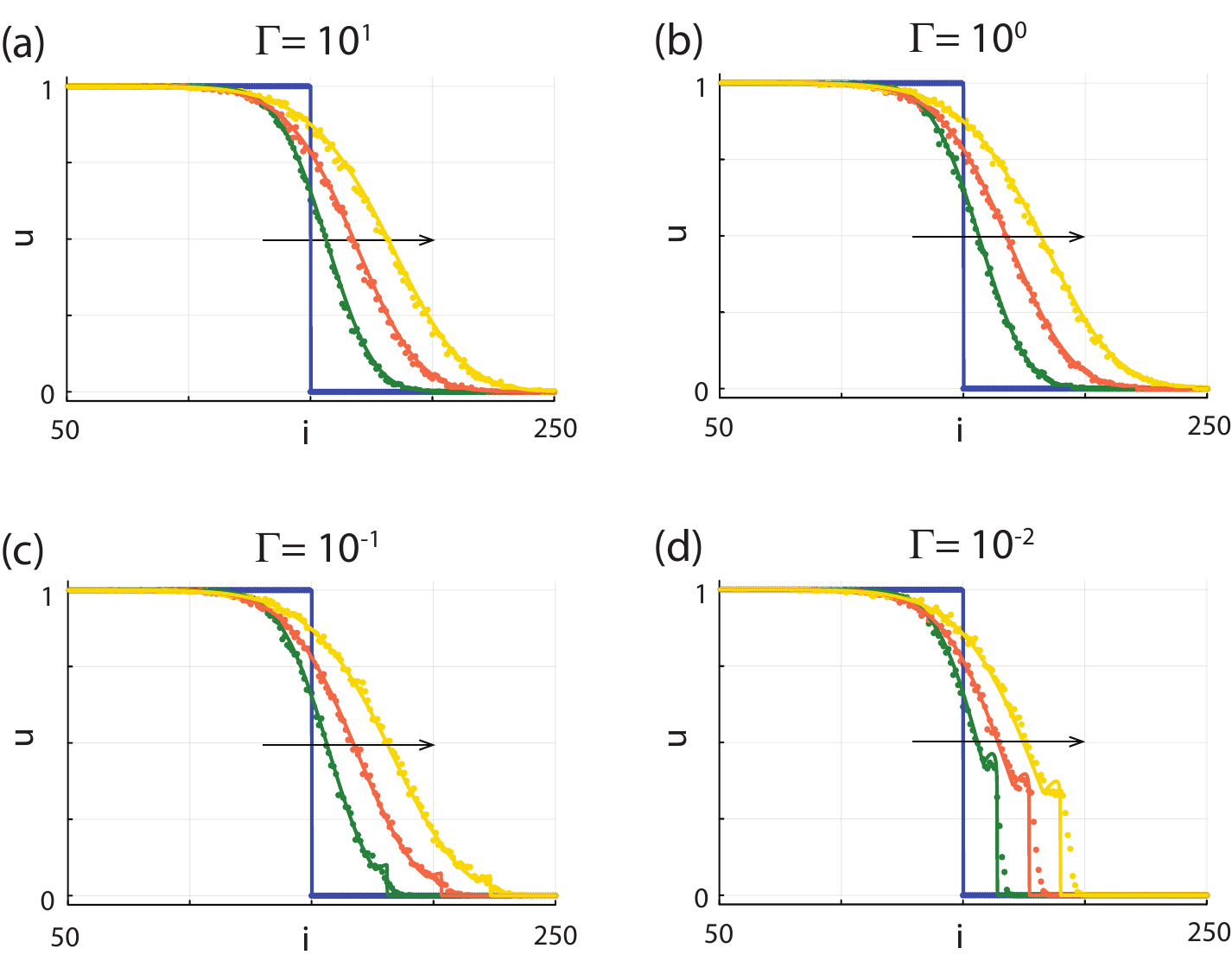}
\renewcommand{\baselinestretch}{1.0}
\caption{Averaged discrete data (dots) superimposed on numerical solutions of Equations \ref{eq:continuousu2}--\ref{eq:continuouss2} (solid).  Each subfigure compares averaged discrete data, constructed using Equation \ref{eq:1Dverticalaverage} with $H=20$, $M = 100$, $P=1$ and $Q=1/500$ with a numerical solution of Equations \ref{eq:continuousu2}--\ref{eq:continuouss2}.  Four sets of solutions are shown for $\Gamma = 10^{1}, 10^{0}, 10^{-1}$ and $10^{-2}$, as indicated.   Within each subfigure a comparison is made at $t=0, 1, 2, 3$ and $4$ days shown in blue, green orange and yellow, respectively, as indicated.   Each day of simulation corresponds to 500 time steps in the discrete model, and snapshots are reported in terms of the $(i,j)$ index of the lattice, which can be re-scaled to give the dimensional coordinates noting that $(x,y) = (i \Delta, j \Delta)$.} \label{F8}
\end{figure}

We now repeat the comparison of averaged discrete data with the solution of  Equations \ref{eq:continuousu2}--\ref{eq:continuouss2} in a one--dimensional Cartesian coordinate system for the same domain, initial conditions and parameter values considered previously in Figure \ref{F4} except now we consider simulations that include proliferation with $Q = 1/500$.  The quality of the continuum--discrete match in Figure \ref{F8} is very good for all values of $\gamma$ considered.  Comparing the solution profiles in Figures \ref{F4} and \ref{F8} shows that the presence of proliferation over a period of four days increases the distance that the population front moves in the positive $x$--direction, just as we demonstrated using the stochastic model in Figure \ref{F7}.  In addition to noting that the numerical solution of   Equations \ref{eq:continuousu2}--\ref{eq:continuouss2}  provides a reasonable match to averaged discrete data, it is important to note that the presence of proliferation in Figure \ref{F8} does not alter the trends established previously in Figure \ref{F4} regarding how $\Gamma$ affects the sharpness of the front, namely that sufficiently large substrate deposition rates leads to smooth--fronted profiles whereas reduced substrate deposition rates leads to sharp--fronted profiles, with the possibility of having a non-monontone shape.

\cbl
Visually comparing results in Figure \ref{F2} and Figure \ref{F7} provides a simple and easy-to-interpret qualitative comparison of the impact of proliferation.  In contrast, comparing results in Figure \ref{F4} and Figure \ref{F8} provides a quantitative comparison of the impact of proliferation since these plots involve identical initial conditions, geometry and visualisation of agent densities over the same time scales.  The only difference is that Figure \ref{F4} involves motility only whereas Figure\ref{F8} involves combined motility and proliferation.  Results in Figure \ref{F8} correspond to biologically--relevant parameter estimates where $Q/P =1/500 \ll 1$ where the mean-field approximation is known to be accurate~\cite{Simpson2010,Baker2010,Simpson2011}.  Other parameter choices $Q/P$ is not sufficiently small lead to situations where the solution of the mean-field model does not match averaged data from the discrete model as explored in~\cite{Simpson2010,Baker2010,Simpson2011}.  Should the reader wish to explore the implications of parameter choices where $Q/P$ is not sufficiently small they may use the open source code on \href{https://github.com/ProfMJSimpson/DiscreteSubstrate}{GitHub}. to make additional continuum--discrete comparisons.  In this work we restrict our attention to biologically relevant parameter estimates where the mean--field approximation is reasonably accurate.
\cb

\section{Conclusion and Outlook}
In this work we have revisited the question of using continuum PDE models to study spatial spreading and invasion of populations of cells.  While many continuum PDE models involve linear diffusion, solutions of these models do not have compact support, and do not replicate clearly defined fronts that are often observed experimentally. Previously, this issue has been addressed by generalising the linear diffusion flux, $\mathbfcal{J} = -D \nabla u$, to a degenerate nonlinear diffusion flux, $\mathbfcal{J} = -D u^n \nabla u$ where $n > 0$.  The motivation for working with degenerate nonlinear diffusion is that the flux vanishes when $u=0$ and the solution of the PDE model has a well--defined sharp front that can match experimental observations~\cite{Vazquez2006,Pattle1959,Barenblatt96,Aronson1985}.  While PDE models with this kind of degenerate nonlinear diffusion flux leads to solutions with well--defined sharp fronts, the biological motivation for these models and a biological interpretation of the exponent $n$ remains unclear.  In this work we have revisited the question of modelling spatial spreading and cellular invasion from the point of view of developing  simple lattice--based discrete model.  In the discrete model we assume that agents produce an external substrate (e.g. biomacromolecules, extracellular matrix) that is deposited locally on the lattice, and the rate of randomly--directed agent migration is taken to be proportional to the density of substrate at each lattice site.  We explicitly incorporate crowding effects in the discrete model by allowing each lattice site to be occupied by, at most, one single agent.  This simple, biologically-motivated mechanism allows us to model collective spreading and invasion with well--defined sharp fronts provided that the rate of substrate deposition is sufficiently small.  Stochastic simulations that mimic the spatial spreading of cells in a two--dimensional circular barrier assay illustrate that our discrete model is capable of replicating key features of the experiment, namely symmetric spreading of the population with a well--defined sharp front at the leading edge of the population.

Coarse--graining the discrete mechanisms leads to a PDE model with a novel flux term that simplifies to linear diffusion in the bulk of the population, and has features similar to a degenerate nonlinear diffusion flux at the leading edge of the population.  Importantly, these features arise within the context of a simple, biologically--motivated discrete mechanism that is capable of replicating sharp--fronted density profiles and our approach does not involve specifying a degenerate nonlinear diffusivity function that is difficult to relate to biological mechanisms.  Numerical solutions of the new PDE model provide us with a computationally efficient, accurate approximation of averaged data from the stochastic model.  Careful examination of the solutions of the PDE indicate that the structure of the leading edge depends upon the rate of substrate deposition.  For sufficiently fast substrate deposition the substrate profile approaches a step function at the leading edge of the spreading population, and the nonlinear PDE model simplifies to the linear diffusion equation.  In contrast, for sufficiently slow substrate deposition the leading edge of the population behaves like a moving boundary problem where the density profile has compact support, and the shape of the density profile at the leading edge can be non-monotone.  The first set of stochastic simulations and coarse--grained PDE models presented in this work focus on conservative populations where cell proliferation and cell death are absent.  To understand how the shape of the front could change when considering a proliferative population  we present a second set of simulations and coarse--grained PDE models that incorporate a minimal proliferation mechanisms.  In this case the coarse--grained PDE model takes the form of a reaction--diffusion model.  We solve the new PDE numerically, using biologically motivated parameter values which show that numerical solutions of the PDE model matches averaged data from the discrete simulations very well, and confirms that sharp--fronted density profiles occur in the presence of proliferation.  In fact, for the biologically motivated parameter values considered in this work, we find that incorporating proliferation leads tends to sharpen the density fronts at the leading edge relative to non-proliferative stochastic simulations.

There are many options for extending the work presented in this study.  One obvious avenue for exploration is to introduce additional details into the discrete model since our approach in this work is to introduce very simple mechanisms only.  An interesting option for further examination would be to generalise the transition probabilities in the following way.  In the current model the transition probability for an agent undergoing a motility event from site $(i,j)$ to site $(i+1,j)$ is proportional to $S_{i,j} \langle U_{i,j} \rangle \left(1 -  \langle U_{i+1,j} \rangle \right)$, which indicates that the transition probability is a linearly increasing function of local substrate density $S_{i,j}$.  An interesting extension would be to generalise the transition probability to be proportional to $g(S_{i,j}) \langle U_{i,j} \rangle \left(1 -  \langle U_{i+1,j} \rangle \right)$, where $0 \le g(S) \le 1$ is smooth function describing how the motility probability for individual agents depends upon the substrate density.  In the context of modelling cell migration it is natural to assume that $g(S)$ is an increasing function.  Taking the continuum limit of the discrete mechanism under these circumstances leads to
\begin{align}
\dfrac{\partial u}{\partial t} =& D \nabla \cdot \left[ g(s) \nabla u + \dfrac{\textrm{d}g(s)}{\textrm{d}s} u\left(1-u\right) \nabla s  \right] + \lambda u(1-u), \label{eq:continuousuextension}\\
\dfrac{\partial s}{\partial t} =& \begin{cases}
                 \gamma u  \quad \textrm{for} \quad s < 1 \label{eq:continuoussextension} \\
                 0  \, \, \, \, \quad \textrm{for}  \quad s  = 1,
                 \end{cases}
\end{align}
which is a generalisation of setting $g(s) = s$.  Returning to our initial discussions in the Introduction, choosing $g(s) = s^n$, for $n > 0$ means that the diffusive flux term in Equations \ref{eq:continuousuextension}--\ref{eq:continuoussextension} is analogous to the flux term in the generalised porous medium equation~\cite{Vazquez2006,Aronson1985,Johnston2023}.  All results presented in this study involves working with the simple choice of $g(s) = s$, however generating and comparing averaged discrete data with numerical solutions of Equations \ref{eq:continuousuextension}--\ref{eq:continuoussextension} would be very interesting to explore how different choices of $g(s)$ might impact the quality of the discrete--continuum match and the shape of the front.  Another extension would be to couple the probability of proliferation in the discrete model to the substrate density.  This would, in effect, introduce a substrate--dependent proliferate rate $\lambda(s)$ into Equation \ref{eq:continuousuextension}.  Again, the question of generating and comparing averaged discrete density data for this generalisation would be interesting and a relatively straightforward extension of the current discrete and continuum modelling frameworks established in this work.

Another extension would be to examine long time travelling wave solutions of Equations \ref{eq:continuousu2}--\ref{eq:continuouss2}~\cite{Murray02}.  In the current work we have limited our examination of this model to relatively short--time simulations of the discrete model and relatively short time numerical solutions of the continuum limit PDE which is relevant when using these models to mimic standard experimental protocols.  Standard experimental protocols examining collective cell migration and proliferation are typically limited to durations of 24 or 48 hours~\cite{Jin2016,Das2015}.  This means that for a typical cell line with a doubling time of 12--24 hours, these standard experimental protocols last for approximately one--to--four times the cell doubling time.  This means that standard experimental protocols are perfectly suited to examine the effects of proliferation that will be evident over these typical timescales.  In our theoretical comparison of averaged discrete data and the solution of the continuum--limit PDE in Figure \ref{F8} is relevant for such typical expeirmental durations since we compare the evolution of the front position over four days for a population with a doubling time of 18 hours, which is just over five times the doubling time. Despite the fact that we have considered numerical solutions of Equations \ref{eq:continuousu2}--\ref{eq:continuouss2} over time scales that are five times the doubling time. it is clear that the numerical solutions in Figure \ref{F8}  have not had sufficient time to approach a constant speed, constant shape travelling wave solution~\cite{Murray02}.  Therefore, taking a more theoretical point of view, it would be mathematically interesting to examine time--dependent numerical solutions of Equations \ref{eq:continuousu2}--\ref{eq:continuouss2} over much longer time scales and study the resulting travelling wave behaviour as $t \to \infty$.  This could be achieved by transforming the time--dependent PDE model into the travelling wave coordinate, $z = x - ct$, where $c$ is the long--time asymptotic speed of the travelling wave solutions.  Properties of the solution of the resulting dynamical system could then be studied in the phase space to provide information about the relationship between parameters in the continuum PDE model and the travelling wave speed $c$ and the shape of the travelling wave profile~\cite{Murray02,ElHachem2022}.   We leave both these potential extensions for future consideration.\\

\noindent
\textbf{Data Accessibility} Open source Julia implementations of all computations are available on GitHub \href{https://github.com/ProfMJSimpson/DiscreteSubstrate}{https://github.com/ProfMJSimpson/DiscreteSubstrate}.\\

\noindent
\textbf{Authors' Contributions} MJS: Conceptualisation, Formal analysis, Investigation, Methodology, Software, Validation, Writing - original draft.  KMM: Conceptualisation, Formal Analysis, Methodology, Software, Validation, Writing - review \& editing.  SWM: Investigation, Writing - review \& editing. PRB: Investigation, Methodology, Writing - review \& editing.\\

\noindent
\textbf{Competing Interests} We declare we have no competing interests.\\

\noindent
\textbf{Funding} MJS and PRB are supported by the Australian Research Council (DP230100025).\\

\noindent
\textbf{Acknowledgements} We thank the Faculty of Science at QUT for providing KMM with a mid-year research fellowship to support this project.

\newpage

\section*{Appendix: Numerical Methods}
Results in Figure \ref{F1} involve generating numerical solutions of
\begin{equation}\label{eq:2DDiffusion}
\dfrac{\partial u}{\partial t} = \dfrac{\partial}{\partial x}\left[ \mathcal{D}(u) \dfrac{\partial u}{\partial x}\right] + \dfrac{\partial}{\partial y}\left[ \mathcal{D}(u) \dfrac{\partial u}{\partial y}\right],
\end{equation}
on a square domain centered at the origin with side length $L$.  To solve Equation \ref{eq:2DDiffusion} we discretise all spatial derivative terms on a uniform square mesh with mesh spacing $h$ so that the mesh point with index $(i,j)$ is associated with location $(-L/2 + (i-1)h, -L/2 + (j-1)h)$.  Applying a standard central difference approximation to the spatial derivative terms in Equation \ref{eq:General} at the central nodes leads to
\begin{align}
& \dfrac{\textrm{d} u_{i,j}}{\textrm{d} t} = \dfrac{1}{2h^2} \left[\left(\mathcal{D}(u_{i,j})+\mathcal{D}(u_{i+1,j}) \right) \left (u_{i+1,j}-u_{i,j}  \right ) - \left(\mathcal{D}(u_{i,j})+\mathcal{D}(u_{i-1,j}) \right) \left (u_{i,j}-u_{i-1,j}  \right ) \right. \notag \\
&+ \left. \left(\mathcal{D}(u_{i,j})+\mathcal{D}(u_{i,j+1}) \right) \left (u_{i,j+1}-u_{i,j}  \right ) - \left(\mathcal{D}(u_{i,j})+\mathcal{D}(u_{i,j-1}) \right) \left (u_{i,j}-u_{i,j-1}  \right ) \right].
\end{align}
This central difference formula is adjusted along the domain boundaries to enforce no-flux boundaries.   When we apply this discretisation to simulate linear diffusion we set $\mathcal{D}(u) = D$, and when simulate nonlinear degenerate diffusion we set $\mathcal{D}(u) = Du^n$.  This system of coupled nonlinear ordinary differential equations is solved using the DifferentialEquation.jl package in Julia, which uses automatic time stepping routines to minimise truncation error.  Results in Figure \ref{F1} are obtained with $h=0.05$, which is sufficiently small to ensure that these numerical results are grid-independent.

Results in the main document include numerical solutions of Equations \ref{eq:continuousu}--\ref{eq:continuouss} in a one--dimensional Cartesian geometry,
\begin{align}
\dfrac{\partial u}{\partial t} =& D \dfrac{\partial }{\partial x} \left[s \dfrac{\partial u}{\partial x} + u(1-u) \dfrac{\partial s}{\partial x}  \right]  + \lambda u (1-u), \label{eq:continuousu1D}\\
\dfrac{\partial s}{\partial t} =& \begin{cases}
                 \gamma u  \quad \textrm{for} \quad s < 1 \label{eq:continuouss1D} \\
                 0  \, \, \, \, \quad \textrm{for}  \quad s  \ge 1,
                 \end{cases}
\end{align}
on $0 < x < L$.  To solve Equations \ref{eq:continuousu1D})--\ref{eq:continuouss1D} we discretise all spatial derivative terms on a uniform mesh with mesh spacing $h$ so that the $i$th mesh point is associated with position $x_i = (i-1)h$.  Applying a standard central difference approximation to the spatial derivative terms in Equation \ref{eq:continuousu1D} gives the following system of coupled nonlinear ordinary differential equations at the $i$th node,
\begin{align}
&\dfrac{\textrm{d} u_i}{\textrm{d} t} = \dfrac{D}{2h^2}  \left[ \left(s_{i+1}+s_{i}\right)\left(u_{i+1}-u_{i}\right)- \left(s_{i-1}+s_{i}\right)\left(u_{i}-u_{i-1}\right)    \right. \label{eq:numericalu1D}  \\
&+\left. \left( u_{i+1}[1-u_{i+1}] + u_{i}[1-u_{i}] \right)\left(s_{i+1}-s_{i}\right)- \left(u_{i-1}[1-u_{i-1}] + u_{i}[1-u_{i}] \right)\left(s_{i}-s_{i-1}\right)    \right] \notag \\
&+ \lambda u_i (1-u_i)  \notag \\
&\dfrac{\textrm{d} s_i}{\textrm{d} t} = \begin{cases}
                 \gamma u_i  \quad \textrm{for} \quad s_i < 1 \label{eq:numericals1D} \\
                 0 \,  \, \, \, \, \quad \textrm{for}  \quad s_i  \ge 1.
                 \end{cases}
\end{align}
The discrete equations for $s$, Equation \ref{eq:numericals1D} holds for all mesh points $i = 1,2,\ldots,I$ because there are no spatial derivative terms in Equation \ref{eq:continuouss1D} and no boundary conditions need to be imposed.  In contrast, the discrete equations for $u$, Equation \ref{eq:numericalu1D} holds only on the interior mesh points $i=2,3,\ldots,I-1$.  Applying no-flux boundary conditions at $i=1$ and $i=I$ means that we impose the constraints $u_1 = u_2$ and $u_{I-1} = u_I$, respectively.  This system of coupled nonlinear ordinary differential equations is solved using the DifferentialEquation.jl package in Julia, which implements automatic time stepping routines to control temporal truncation error.  All numerical results in this work correspond to $h=0.1$, which is sufficiently small to ensure that our numerical results are grid-independent for the problems that we consider. Open source Julia code to solve Equations \ref{eq:numericalu1D}--\ref{eq:numericals1D} is available on \href{https://github.com/ProfMJSimpson/DiscreteSubstrate}{GitHub}.  \cbl Results in this study are obtained using a 5-4th order Runge-Kutta method within the Tsit5 algorithm within the DifferentialEquations.jl package. \cb

\end{document}